\documentclass{INTERSPEECH2023}

\interspeechcameraready

\title{An Efficient Speech Separation Network Based on Recurrent Fusion Dilated Convolution and Channel Attention}
\name{Junyu Wang}
\address{
  School of Electronic Information, Sichuan University, China}
\email{junyu\_wang@stu.scu.edu.cn}

\begin{document}

\maketitle
 
\begin{abstract}
We present an efficient speech separation neural network, ARFDCN, which combines dilated convolutions, multi-scale fusion (MSF), and channel attention to overcome the limited receptive field of convolution-based networks and the high computational cost of transformer-based networks. The suggested network architecture is encoder-decoder based. By using dilated convolutions with gradually increasing dilation value to learn local and global features and fusing them at adjacent stages, the model can learn rich feature content. Meanwhile, by adding channel attention modules to the network, the model can extract channel weights, learn more important features, and thus improve its expressive power and robustness. Experimental results indicate that the model achieves a decent balance between performance and computational efficiency, making it a promising alternative to current mainstream models for practical applications.
\end{abstract}

\section{Introduction}

The objective of speech separation is to separate the required sound from surrounding noise such as environmental noise and other voices. Depending on the number of channels or microphones, speech separation tasks are frequently divided into multi-channel and single-channel. Because of the challenges in encoding multi-speaker mixture speech signals into a completely separable feature space \cite{interspeech1}, speech separation remains an extremely difficult research subject, particularly in the case of single-channel. We focus on single-channel speech separation.

Numerous attempts have been undertaken throughout the years to tackle this issue, despite the fact that single-channel speech separation presents numerous difficulties. Before the age of deep learning, a number of classical approaches were used to this problem, including non-negative matrix factorization (NMF) \cite{interspeech2, interspeech3}, computational auditory scene analysis (CASA) \cite{interspeech5}, and support vector machine (SVM) \cite{interspeech4}. However, these models frequently struggle to represent the nonlinear connection between speech data, which severely restricts their practical utilization. With the emergence of deep learning approaches in other disciplines \cite{interspeechcbam, interspeechresnet}, researchers have begun to employ massive amounts of data to construct models capable of separating mixed speech from unknown speakers; as a result, the performance of this task has been significantly enhanced. Currently, single-channel speech separation approaches are primarily classified into two categories: end-to-end time domain and time-frequency (T-F) domain. In the T-F domain, the input signal is first transformed into T-F features using the short-time Fourier transform (STFT), followed by the hadamard product of the amplitude spectrum of the mixed signal and the predicted masking value to obtain the estimated amplitude spectrum of the target signal. Lastly, the enhanced amplitude spectrum and the original phase spectrum \cite{interspeech8, interspeech10, interspeech11, interspeech12} are subjected to the inverse Fourier transform (ISTFT) to separate speech.

However, due to the difficulty in reconstructing the waveform phase, the predicted source waveform is typically synthesized by mixing the initial phase, reducing the performance ceiling of the method. To overcome this issue, the paper \cite{interspeech13} introduced the time-domain speech separation approach, which employs the codec framework based on convolutional neural networks to model the mixed waveform directly, and the separation section is comprised of multiple convolutional filters, which has achieved good results. Numerous subsequent works \cite{interspeech16, interspeech14, interspeech18, interspeech19} are devoted to constructing better separators on TASNET to better enhance speech separation performance. 

Among these time-domain approaches, convolution-based multi-scale fusion (MSF) methods \cite{interspeech20, interspeech21} and transformer-based methods \cite{interspeech25, interspeech37} have achieved excellent performance in speech separation tasks by learning features at different time scales. However, transformer-based models typically require enormous computational cost and have long training time, making them less suitable for practical applications. Convolution-based models often encounter the issue of limited receptive field, which causes them to fall short of reaching the theoretical upper limit of the model. Additionally, we have observed that few methods in speech separation focus on channel attention, which may lead to difficulty capturing the channel correlations between input features, thus limiting performance.

Based on these findings, we present an ARFDCN that incorporates MSF, dilated convolutions, and channel attention, which makes full use of the time dependence of speech signals by combining exponentially growing dilated convolution \cite{interspeech22, interspeech23} with MSF to ensure that the network has dynamic receptive fields, so as to achieve excellent separation performance. Moreover, channel attention is added after each MSF block, allowing the network to focus on specific regions and thus enhance the ability of feature representation and discrimination of the network. Because of the application of dilated convolutions, MSF, and channel attention, the features derived from the network during the speech separation stage are more comprehensive and can make better use of the information between contexts.

\begin{figure*}[htbp]
\centering
\includegraphics[width=0.8\linewidth]{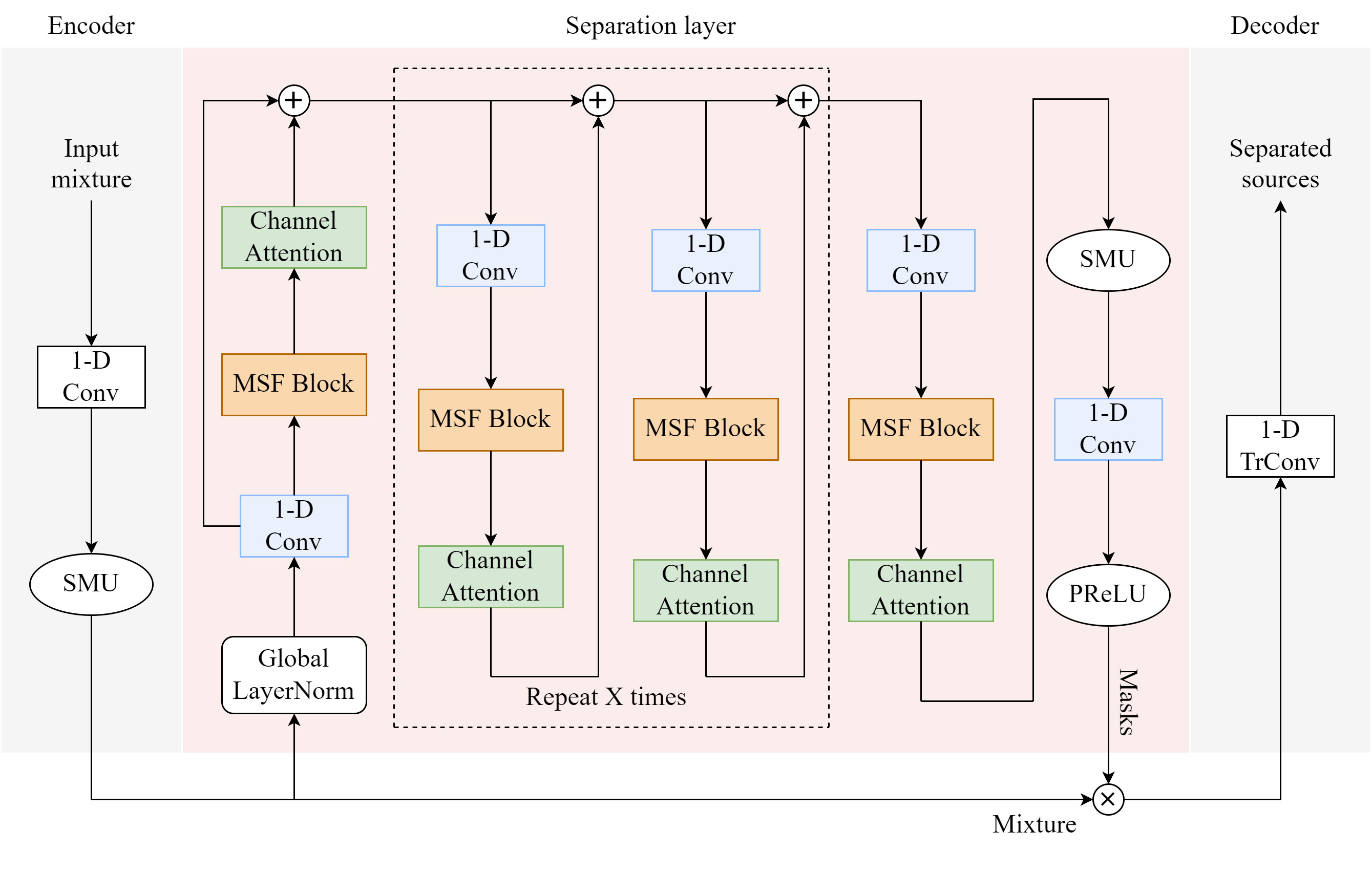}
\captionsetup{justification=raggedright}
\caption{System flowchart of ARFDCN.}
\label{fig1}%文中引用该图片代号
\end{figure*}

\section{ARFDCN}

The presented ARFDCN, like Conv-Tasnet \cite{interspeech14}, involves three phases: encoder, separation layer, and decoder. First, the encoder converts the blended speech waveform into corresponding features in the separable embedding feature space. After the features have been sent, a mask for each source is created by the separation layer. Finally, by transforming the mask features, the decoder reconstructs the waveforms of each source \cite{interspeech25}. The flowchart of the model is represented in Figure~\ref{fig1}. 

\subsection{Encoder}

The input mixing speech is separated into overlapping segments of length $H$ in this module, denoted by ${E}_{s} \in R^{1 \times H}$, where $s=1,...,N$ indicates the segment index and $N$ is the total number of input segments. 
By using a trainable 1-D convolution operation and SMU activation \cite{interspeechsmu}, the following ${E}_{s}$ is encoded into a high-dimensional mixed feature ${X} \in R^{C \times L}$:
\begin{equation}
{X} = SMU(Conv1D({x}))
\end{equation}
where $C$ represents the number of channels and $L$ is the signal length generated by 1-D convolution.

\subsection{Separation layer} 

The separation layer is inspired by \cite{interspeechcbam, interspeech21}. It is predominantly made up of channel attention, variable dilated convolutions and MSF. After the output of the encoder is delivered to the separation layer, it first undergoes global layer normalization (GLN) \cite{interspeech14} to boost network convergence speed, followed by 1-D convolution operation with \(P\) output channels. After the convolution output, it passes through multiple MSF blocks and channel attention modules in turn.

\begin{figure}[t]
\centering
\includegraphics[width=0.99\linewidth]{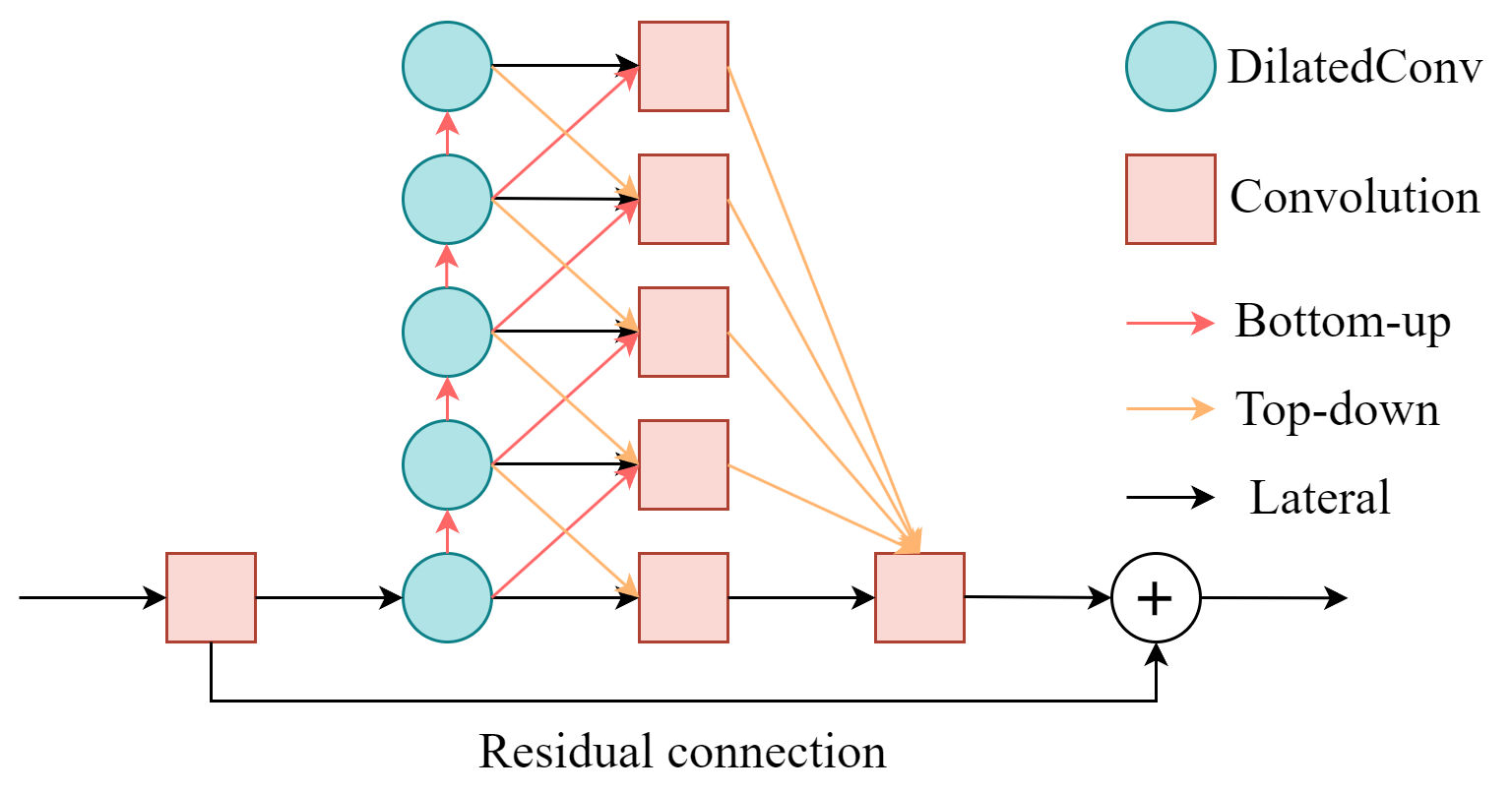}
\captionsetup{justification=raggedright}
\caption{\quad Multi-scale information fusion block.}
\label{fig2}%文中引用该图片代号
\end{figure}

The specific structure of the MSF block is illustrated in Figure~\ref{fig2}. In each MSF block, the input flows from bottom to top, obtaining different receptive field information at different stages, with a total of $J$ stages. The lower-level dilated convolution is used to extract local detail features, while the higher-level dilated convolution is used to extract global semantic features. GLN is inserted between each dilated convolution operation to ensure more stable output data. In addition, to allow the higher-level dilated convolution layer to obtain a sufficiently large receptive field, we use dilation values that increase exponentially in powers of two from bottom to top. By fusing information from these different scales, the model can learn comprehensive features and improve its performance. Each convolution layer in the MSF block is followed by GLN and PReLU function, which helps to improve the expressiveness and robustness of the model. Finally, the residual connection is introduced between the input and output of the MSF block to enhance the representational capacity of the model and alleviate the gradient vanishing issue. Compared to the widely used dual-path network for learning local and global features, the MSF block has a shorter computation time and is more suitable for practical applications.
 
\begin{figure}[t]
  \centering
  \includegraphics[width=\linewidth]{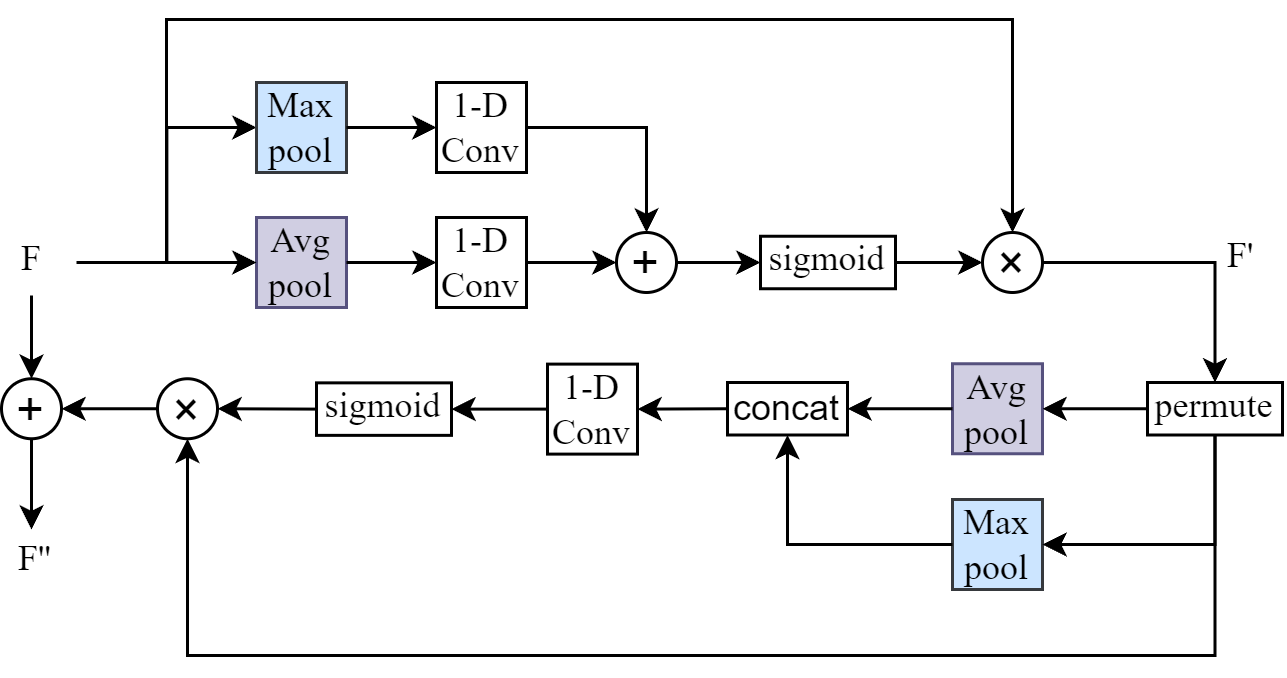}
  \caption{Channel attention module.}
  \label{fig3}
\end{figure}

To further improve the generalization ability of the model and make it more focused on important features, channel attention is added after each MSF block, as shown in Figure~\ref{fig3}. First, maximum pooling and average pooling are applied to the inputs along the last dimension, respectively. Then, a 1-D convolution with a kernel size of 5 is applied to generate a weight vector for each channel. The two weight vectors are added and multiplied with the input feature to weigh each channel and generate a new feature map \(F'\). Next, maximum pooling and average pooling are applied to the new feature vector along its channel dimension, respectively. A 1-D convolution with a kernel size of 21 is applied to generate a weight vector for the spatial position, and the weight vector is multiplied with the input to generate the final feature. Additionally, to alleviate gradient vanishing and improve model stability, residual connection is incorporated into the channel attention module. The formula for channel attention is as follows:
\begin{align}
F' &= \phi (h^{5}(Avp(F)) + h^{5}(Map(F))) * F \\
F'' = \phi (&h^{21}(concat[Avp(F'),Map(F')])) * F' + F
\end{align}
where $Avp(\cdot)$ and $Map(\cdot)$ represent average pooling and max pooling operations, respectively. $h^{5}$ represents a 1-D convolution operation with a kernel size of 5, input channel of 1, and output channel of 1.
$\phi(\cdot)$ represents a sigmoid activation operation. $concat[\cdot]$ represents concatenation along the channel dimension. $h^{21}$ represents a 1-D convolution operation with a kernel size of 21, input channel of 2, and output channel of 1.

\begin{figure}[htbp]
  \centering
  \includegraphics[width=\linewidth]{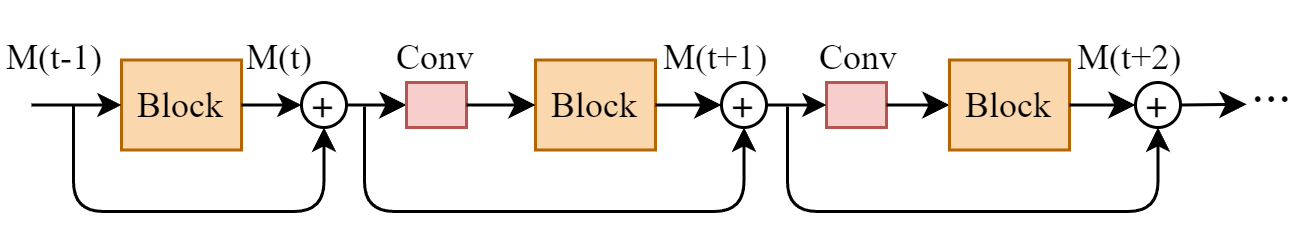}
  \caption{Integration of MSF blocks.}
  \label{fig4}
\end{figure}

Moreover, to better fuse information among MSF blocks and prevent gradient vanishing. Before delivering to the subsequent block, we integrate the output of the current block with the output of all previous blocks and the input of the initial MSF block, and add a 1-D convolution before each input of the MSF block. Each convolution layer is followed by GLN and SMU activation function, as shown in Figure~\ref{fig4}. This approach is described as follows: 
\begin{equation}
M(t+1) = f(c(M(t)+M(t-1)+...+e))
\end{equation}
where $f(\cdot)$ represents the operation of passing through an MSF block, $c(\cdot)$ represents the 1-D convolution operation with GLN and SMU activation, $M(t)$ represents the output of the MSF block at time step $t$, and $e$ represents the input of the initial MSF block. 

The output of the last MSF block is passed through a 1-D convolution operation with PReLU activation to produce a mask for each source.

\subsection{Decoder}
The 1-D transposed convolution process of the decoder employs the same stride and convolution kernel as the encoder. The input of the decoder is obtained by element-wise multiplication between the output of the encoder X and the mask ${D}_{i}$ of the i-th source. The transformation of the decoder can be represented as follows:
\begin{equation}
{Y}_{i} = ConvTranspose1d({D}_{i}\ast {X})
\end{equation}
where ${Y}_{i} \in R^{1 \times T}$ is the final waveform of the i-th speaker.

\section{Experimental procedures}
\label{sec:guidelines}

\subsection{Dataset}
Using three publicly accessible English datasets, WSJ0-2Mix \cite{interspeech8}, Libri2mix \cite{interspeechlibrimix}, and WHAM! \cite{interspeechwham}, we examine the separation performance of the ARFDCN model. The voice source for the WSJ0-2Mix is the WSJ0 \cite{interspeechwsj0}. The si\_tr\_s file generated training data for 30 hours and validation data for 10 hours. The si\_dt\_05 and si\_et\_05 files created 5 hours of test data. The WHAM! is a noisy speech-mixed dataset based on the WSJ0-2Mix. This dataset is created by adding a different ambient noise sample to each of the two speakers' mixed speech from the WSJ0-2Mix. In this experiment, it is required to not only separate the speaker signal but also reduce the background noise. This task is more difficult than separating the WSJ0-2Mix. The Libri2Mix is created using data from the LibriSpeech corpus \cite{interspeechlibrispeech}. In the LibriSpeech, different speaker signals are arbitrarily extracted and combined using the uniform sampling loudness unit of relative full scale (LUFS) \cite{interspeechlufs} between -25 and -33 dB. Random noise samples are added to the speech mixture, uniformly distributed with loudness between -38 and -30 LUFS. The training set is the subset train-100. All speech signals are resampled at a frequency of 8kHz.

\subsection{Experiment configurations}
Encoder and decoder are configured with 512 channels, kernel size of 21, and stride size of 10. The number of times X and channels \(P\) in the separation layer are set to 7 and 512. The value of $J$ in the MSF block is set to 5, and the dilation values for the dilated convolutions are set from low to high as \{1, 2, 4, 8, 16\}, with a kernel size of 5 and a stride size of 2. 

The model is trained for 200 epochs. In WSJ0-2Mix and WHAM!, the training speech segment length is set to 4-second, whereas in Libri2Mix, the training speech segment length is set to 3-second. We utilize AdamW \cite{interspeechadamw} as the optimizer and set the initial learning rate to $1 \times 10^{-3}$ during training. If the validation accuracy does not increase for two successive epochs, the learning rate will be reduced to 0.9 times. 

To maximize the signal fidelity of the model, this paper takes the scale-invariant source-to-distortion ratio (SI-SDR) \cite{interspeech34, interspeech35} as the training target. It can be described as:

\begin{equation}
  s_{target}=\frac{\langle\tilde y, y\rangle y}{\lVert{y}\rVert^2}
  \label{eq_si-snr_1}
\end{equation}
\begin{equation}
  e_{noise}=\tilde y-s_{target}
  \label{eq_si-snr_2}
\end{equation}
\begin{equation}
  SI-SDR=10\log_{10}{\frac{\lVert{s_{target}}\rVert^2}{\lVert{e_{noise}}\rVert^2}}
  \label{eq_si-snr_3}
\end{equation}
where $\tilde y$ and $y$ represent estimated and clean sources, respectively. Prior to calculation, both are normalized to zero-mean. 

\section{Results}

\subsection{Comparison with other models}

We begin by presenting the results for the WSJ0-2Mix dataset. Table~\ref{tab1} compares ARFDCN with multiple prevalent models on the WSJ0-2Mix, where the ``Time'' metric is computed by processing a 4-second audio sample with a sample rate of 8kHz on Intel(R) Core(TM) i9-12900KF CPU @ 3.19GHz, averaged over 100 attempts. The results indicate that Sepformer achieves only modest improvement over ARFDCN, despite significantly increasing the model size and reasoning time. 

\begin{table}[h]
  \caption{Evaluation of different speech separation models utilizing Si-SDRi and SDRi on WSJ0-2Mix}
  \label{tab1}
  \centering
  \tabcolsep=0.07cm
  \begin{tabular*}{\hsize}{@{}@{\extracolsep{\fill}}ccccc@{}}
    \hline
    \textbf{Model}  &\textbf{SI-SDRi}  & \textbf{SDRi}    & \textbf{Time(s)}  & \textbf{Para.(M)} \\
    \hline
    DPCL++ \cite{interspeech8}     & 10.8   & 11.2  & 0.47  & 13.6        \\
		uPIT-BLSTM-ST \cite{interspeech11}    & 9.8   & 10.0   & 0.97   & 92.7   \\
		Chimera++ \cite{interspeech12}    & 11.5   & 11.8   & 0.83   & 32.9    \\
		BLSTM-TasNet \cite{interspeech13}    & 13.2   & 13.6   & 3.66   & 23.6 \\
		Conv-TasNet \cite{interspeech14}    & 15.3   & 15.6  & 0.53   & 5.6    \\
		MSGT-TasNet \cite{interspeech18}    & 17.0   & 17.3   & 6.99   & 66.8   \\
		SuDoRM-RF 1.0x \cite{interspeech20}   & 17.0   & 17.3   & 0.67  & 2.66 \\
		DualPathRNN \cite{interspeech19}   & 18.8   & 19.0    & 5.88   & 2.63 \\
		DPTNet \cite{interspeech25}   & 20.2   & 20.6    & 8.41   & 2.69\\
            Gated DPRNN \cite{interspeech36}    & 20.1   & - 
        & 9.25  & 7.5\\
		A-FRCNN-16 \cite{interspeech21}   & 18.3   & 18.6   & 1.91  & 6.1 \\ 
		Sepformer \cite{interspeech37}   & 20.4   & 20.5  & 6.58  & 26.0 \\
    \hline
    \textbf{ARFDCN}    & \textbf{20.3}   & \textbf{20.5}    & \textbf{1.81}   & 6.14   \\               
    \hline
  \end{tabular*}
\end{table}

To demonstrate the generalizability of our method, relevant experiments are conducted using the WHAM! dataset. Compared with the data in the WSJ0-2Mix, random noise is added to make speech separation more difficult. Table~\ref{tab2} presents the findings of the ARFDCN and 10 contrasted methods on WHAM!. From these results, it can be seen that the ARFDCN and Gated DPRNN achieve almost identical performance, with significantly improved computational efficiency and reduced model size.

\begin{table}[h]
	\centering
	\caption{Performance of different models on WHAM!}
        \label{tab2}
	\begin{tabular}{cccc}
		\toprule 
		\textbf{Model}  &\textbf{SI-SDRi}  & \textbf{SDRi}  & \textbf{Para.(M)}  \\
		\midrule  %添加表格中横线
		Chimera++ \cite{interspeech12}   & 10.0   & -    & 32.9      \\
		BLSTM-TasNet \cite{interspeech13}   & 9.8   & -    & 23.6   \\
		Conv-TasNet \cite{interspeech14}   & 12.7   & -    & 5.6  \\
		MGST-TasNet \cite{interspeech18}  & 13.1   & -   & 66.8   \\
		SuDoRM-RF 1.0x \cite{interspeech20}  & 12.9   & 13.3    & 2.66   \\
		DualPathRNN \cite{interspeech19}   & 13.7   & 14.1    & 2.63   \\
		DPTNet \cite{interspeech25}    & 14.9   & 15.3    & 2.69   \\
		Gated DPRNN \cite{interspeech36}   & 15.2   & -    & 7.5   \\
		A-FRCNN-16 \cite{interspeech21}    & 14.5   & 14.8    & 6.1   \\
            Sepformer \cite{interspeech37}   & 14.4   & 15.0    & 26.0  \\
		\midrule
		\textbf{ARFDCN}    & \textbf{15.2}   & \textbf{15.5}    & 6.14  \\
		\bottomrule %添加表格底部粗线
	\end{tabular}
\end{table}

Lastly, we verify the performance of ARFDCN on the Libri2Mix dataset. The dataset combination is more challenging to separate than the WSJ0-2Mix, and random noise has been included to get it closer to real application. Table~\ref{tab3} presents the outcomes of the ARFDCN and 8 compared methods on the Libri2Mix. As can be shown, the suggested method ARFDCN is superior to the compared methods, which further demonstrates its effectiveness.

\begin{table}[t]
	\centering
	\caption{Performance of different models on Libri2Mix}
        \label{tab3}
	\begin{tabular}{cccc}
		\toprule 
		\textbf{Model}  &\textbf{SI-SDRi}  & \textbf{SDRi}  & \textbf{Para.(M)}  \\
		\midrule  %添加表格中横线
		Chimera++ \cite{interspeech12}   & 6.3   & 7.0    & 32.9      \\
		BLSTM-TasNet \cite{interspeech13}   & 7.9   & 8.7    & 23.6   \\
		Conv-TasNet \cite{interspeech14}   & 12.2   & 12.7    & 5.6  \\
            SuDoRM-RF 1.0x \cite{interspeech20}  & 13.5   & 14.0    & 2.66   \\
		DualPathRNN \cite{interspeech19}   & 16.1   & 16.6    & 2.63   \\
		DPTNet \cite{interspeech25}    & 16.7   & 17.1    & 2.69   \\
		A-FRCNN-16 \cite{interspeech21}    & 16.7   & 17.2    & 6.1   \\
            Sepformer \cite{interspeech37}   & 16.5   & 17.0    & 26.0  \\
		\midrule
		\textbf{ARFDCN}    & \textbf{17.1}   & \textbf{17.6}    & 6.14   \\
		\bottomrule %添加表格底部粗线
	\end{tabular}
\end{table}

\subsection{Ablation study}

We design four experiments to demonstrate the effects of different modules on the performance, conducted on the WSJ0-2Mix dataset. RFCN removes the channel attention module from ARFDCN and sets all dilation values in the MSF block to 1. RFDCN removes only the channel attention module from ARFDCN. ARFCN sets all dilation values in the MSF block to 1 while keeping the channel attention module. Comparing RFDCN and RFCN shows that combining dilated convolutions with MSF is more beneficial for expanding the receptive field and extracting global information. Moreover, Table~\ref{tab4} shows that ARFCN performs better than RFCN, and ARFDCN performs better than RFDCN, demonstrating the effectiveness of channel attention in improving model performance by focusing on important features.

\begin{table}[h]
	\centering
	\caption{Ablation study results on WSJ0-2Mix}
        \label{tab4}
	\begin{tabular}{cccc}
		\toprule 
		\makebox[0.12\textwidth][c]{\textbf{Model}}  & \makebox[0.07\textwidth][c]{\textbf{SI-SDRi}}  & \makebox[0.07\textwidth][c]{\textbf{SDRi}}  & \makebox[0.07\textwidth][c]{\textbf{Time(s)}} \\
		\midrule  %添加表格中横线
		RFCN   & 18.9   & 19.2   & 1.83  \\
		RFDCN    & 19.6   & 19.9  & 1.77 \\
		ARFCN    & 19.7   & 20.0   & 1.87  \\
            ARFDCN  & \textbf{20.3}   & \textbf{20.5}   & 1.81  \\
		\bottomrule %添加表格底部粗线
	\end{tabular}
\end{table}

\section{Conclusion}

This paper presents ARFDCN, a single-channel speech separation neural network that combines dilated convolutions and MSF to learn micro and macro features at different stages and fuses them to obtain large receptive fields. Moreover, incorporating multiple channel attention modules enables the model to focus more on significant features, improving its expressive power and generalization performance. Experiments on three datasets exhibit that the network reaches a decent balance between computational cost and performance.  Future research will focus on the speech separation of multiple speakers under reverberation conditions and noisy environments to better serve society.

\bibliographystyle{IEEEtran}
\bibliography{mybib}

\end{document}